\documentclass[preprint]{revtex4}
\input epsf
\usepackage{amsmath,amssymb}
\usepackage{graphicx}

\draft

\begin{document}
\titlepage
\title{Palatini formulation of modified gravity with squared scalar curvature}
\author{Xinhe Meng$^{1,2}$ \footnote{mengxh@public.tpt.tj.cn}
 \ \ Peng Wang$^1$ \footnote{pewang@eyou.com}
} \affiliation{1.  Department of Physics, Nankai University,
Tianjin 300071, P.R.China \\2. Institute of Theoretical Physics,
CAS, Beijing 100080, P.R.China}

\begin{abstract}
In this paper we derive the Modified Friedmann equation in the
Palatini formulation of $R^2$ gravity. Then we use it to discuss
the problem of whether in Palatini formulation a $R^2$ term can
drive an inflation. We show that the Palatini formulation of $R^2$
gravity cannot lead to gravity-driven inflation. If considering no
zero radiation and matter energy densities, we show that only
under rather restrictive assumption about the radiation and matter
energy densities there will be a mild power-law inflation $a\sim
t^2$.
\end{abstract}

\maketitle

\textbf{1. Introduction}

The expansion of our universe is currently in an accelerating
phase now seems well-established \cite{Perlmutter}. But now the
mechanism responsible for this is not very clear. Many authors
introduce a mysterious cosmic fluid called dark energy to explain
this. See Ref.\cite{Carroll2} for a review and Ref.\cite{Dark} for
some recent models. On the other hand, some authors suggested that
maybe there does not exist such a mysterious dark energy, but the
observed cosmic acceleration is a signal of our first real lack of
understanding of gravitational physics \cite{Lue, Carroll}. An
example is the braneworld theory of Dvali et al. \cite{Dvali}.
Recently, some authors proposed to add a $1/R$ term in the
Einstein-Hilbert action to modify the General Relativity (GR)
\cite{Carroll, Capozziello}. It is interesting that such a term
may be predicted by string/M-theory \cite{Odintsov3}. In the
metric formulation, this additional term will give fourth order
field equations. It has been shown that this additional term can
give accelerating expansion without dark energy \cite{Carroll}. In
this framework, Dick \cite{Dick} considered the problem of weak
field approximation. Soussa and Woodard \cite{Woodard} have
considered the gravitational response to a diffuse source.

Based on this modified action, Vollick \cite{Vollick} has used
Palatini variational principle to derive the field equations. In
the Palatini formulation, instead of varying the action only with
respect to the metric, one views the metric and connection as
independent field variables and vary the action with respect to
them independently. This would give second order field equations.
For the original Einstein-Hilbert action, this approach gives the
same field equations as the metric variation. For a more general
action, those two formulations are inequivalent, they will lead to
different field equations and thus describe different physics
\cite{Volovich}. Flanagan \cite{Flanagan} derived the equivalent
scalar-tensor description of the Palatini formulation. In
Ref.\cite{Dolgov}, Dolgov and Kawasaki have argued that the fourth
order field equations in metric formulation suffer serious
instability problem. If this is indeed the case, the Palatini
formulation appears even more appealing, because the second order
field equations in Palatini formulation are free of this sort of
instability \cite{Wang}. Furthermore, Chiba \cite{Chiba} has
argued that the theory derived using metric variation is in
conflict to the solar system experiments while in the Palatini
formulation, the model seems to be compatible with solar system
experiment \cite{olmo}. Moreover, a convincing motivation to take
the Palatini formalism seriously is that the Modified Friedamnn
(MF) equation following from it fit the SNe Ia data at an
acceptable level \cite{Wang}.

On the other end of cosmic evolution time, the very early stage,
it is now generally believed that the universe also undergos an
acceleration phase called inflation. The mechanism driven
inflation is also not very clear now. The most popular explanation
is that inflation is driven by some inflaton field \cite{Liddle}.
Also, some authors suggest that modified gravity could be
responsible for inflation \cite{Starobinskii, Odintsov}. Revealing
the mechanisms of inflation is also one of the most important
problems in modern cosmology.

As originally proposed by Carroll et al. \cite{Carroll} and later
implemented by Nojiri and Odintsov \cite{Odintsov}, adding
correction term $R^m$ with $m>0$ in addition to the $1/R$ term may
explain both the early time inflation and current acceleration
without inflaton and dark energy. Furthermore, Nojiri and Odintsov
\cite{Odintsov} showed that adding a $R^m$ term can avoid the
above mentioned instability when considering the theory in metric
formulation. In this paper, we will show that in the Palatini
formulation, the $R^2$ term cannot lead to a gravity-driven
inflation, in opposite to the conclusion when considering the
theory in metric formulation \cite{Starobinskii}.

This paper is arranged as follows: in Sec.2 we review the
framework of deriving field equations and Modified Friedmann (MF)
equations in Palatini formulation; in Sec.3 we discuss the $R^2$
gravity in Palatini formulation and inflation in this model;
Sec.4 is devoted to conclusions and discussions.

\textbf{2. Deriving the Modified Friedmann equation in Palatini
formulation}

First, we briefly review deriving field equations from a
generalized Einstein-Hilbert action of the form $L(R)$ by using
Palatini variational principle. See Refs.\cite{Wang} for details.

The field equations follow from the variation in Palatini approach
of the generalized Einstein-Hilbert action
\begin{equation}
S=-\frac{1}{2\kappa^2}\int{d^4x\sqrt{-g}L(R)}+S_M\ ,\label{action}
\end{equation}
where $\kappa^2 =8\pi G$, $L$ is a function of the scalar
curvature $R$ and $S_M$ is the matter action.

Varying with respect to $g_{\mu\nu}$ gives
\begin{equation}
L'(R)R_{\mu\nu}-\frac{1}{2}L(R)g_{\mu\nu}=\kappa^2 T_{\mu\nu}\
,\label{2.2}
\end{equation}
where a prime denotes differentiation with respect to $R$ and
$T_{\mu\nu}$ is the energy-momentum tensor given by
\begin{equation}
T_{\mu\nu}=-\frac{2}{\sqrt{-g}}\frac{\delta S_M}{\delta
g^{\mu\nu}}\ .\label{2.3}
\end{equation}
We assume the universe contains dust and radiation, thus
$T^{\mu}_{\nu}=\{-\rho_m-\rho_r,p_r,p_r,p_r\}$ where $\rho_m$ and
$\rho_r$ are the energy densities for dust and radiation
respectively, $p_r$ is the pressure of the radiation. Note that
$T=g^{\mu\nu}T_{\mu\nu}=-\rho_m$ because of the relation
$p_r=\rho_r/3$.

In the Palatini formulation, the connection is not associated with
$g_{\mu\nu}$, but with $h_{\mu\nu}\equiv L'(R)g_{\mu\nu}$, which
is known from varying the action with respect to $\Gamma
^{\lambda}_{\mu\nu}$. Thus the Christoffel symbol with respect to
$h_{\mu\nu}$ is given by
\begin{equation}
\Gamma
^{\lambda}_{\mu\nu}=\{^{\lambda}_{\mu\nu}\}_g+\frac{1}{2L'}[2\delta
^{\lambda}_{(\mu}\partial
_{\nu)}L'-g_{\mu\nu}g^{\lambda\sigma}\partial _{\sigma}L']\
,\label{Christoffel}
\end{equation}
where the subscript $g$ signifies that this is the Christoffel
symbol with respect to the metric $g_{\mu\nu}$.

The Ricci curvature tensor is given by
\begin{eqnarray}
R_{\mu\nu}=R_{\mu\nu}(g)+\frac{3}{2}(L')^{-2}\bar{\nabla}
_{\mu}L'\bar{\nabla} _{\nu}L' -(L')^{-1}\bar{\nabla}
_{\mu}\bar{\nabla}
_{\nu}L'-\frac{1}{2}(L')^{-1}g_{\mu\nu}\bar{\nabla}
_{\sigma}\bar{\nabla} ^{\sigma}L'\ ,\label{Ricci}
\end{eqnarray}
and
\begin{equation}
R=R(g)-3(L')^{-1}\bar{\nabla} _{\mu}\bar{\nabla} ^{\mu}
L'+\frac{3}{2}(L')^{-2}\bar{\nabla}_{\mu}L'\bar{\nabla}^{\mu}L'\
,\label{scalar}
\end{equation}
where $R_{\mu\nu}(g)$ is the Ricci tensor with respect to
$g_{\mu\nu}$, $R=g^{\mu\nu}R_{\mu\nu}$ and $\bar{\nabla}$ is the
connection associated to $g_{\mu\nu}$. Note by contracting
(\ref{2.2}), we get:
\begin{equation}
L'(R)R-2L(R)=\kappa^2 T\ .\label{R(T)}
\end{equation}
Assume we can solve $R$ as a function of $T$ from (\ref{R(T)}).
Thus (\ref{Ricci}), (\ref{scalar}) do define the Ricci tensor with
respect to $h_{\mu\nu}$.

Then let's  derive the MF equation in Palatini formulation. Let us
work with the Robertson-Walker metric,
\begin{equation}
ds^2=-dt^2+a(t)^2(dx^2+dy^2+dz^2)\ .\label{metric}
\end{equation}
Note that we only consider a flat metric, which is favored by
present observations \cite{Perlmutter}.

From (\ref{metric}), (\ref{Ricci}), we can get the non-vanishing
components of the Ricci tensor:
\begin{equation}
R_{00}=-3\frac{\ddot{a}}{a}+\frac{3}{2}(L')^{-2}(\partial_0{L'})^2-\frac{3}{2}(L')^{-1}\bar{\nabla}_0\bar{\nabla}_0L'\
,\label{R00}
\end{equation}
\begin{eqnarray}
R_{ij}=[a\ddot{a}+2\dot{a}^2+(L')^{-1}\{^0_{ij}\}_g\partial_0L'
+\frac{a^2}{2}(L')^{-1}\bar{\nabla}_0\bar{\nabla}_0L']\delta_{ij}\
.\label{ij}
\end{eqnarray}

Substituting equations (\ref{R00}) and (\ref{ij}) into the field
equations (\ref{2.2}), we can get
\begin{equation}
6H^2+3H(L')^{-1}\partial_0L'+\frac{3}{2}(L')^{-2}(\partial_0L')^2=\frac{\kappa^2
(\rho+3p)+L}{L'}\ ,\label{aa}
\end{equation}
where $H\equiv \dot{a}/a$ is the Hubble parameter, $\rho$ and $p$
are the total energy density and total pressure respectively.
Assume that we can solve $R$ in term of $T$ from Eq.(\ref{R(T)}),
substitute it into the expressions for $L'$ and $\partial_0L'$, we
can get the MF equation.

In this paper, we will consider the Palatini formulation of the
modified action with a $R^2$ term,
\begin{equation}
L=R+\frac{R^2}{3\beta}\ ,\label{R2}
\end{equation}
where $\beta$ is a constant having the dimension of $(mass)^2$.
This action has been studied by Starobinsky in metric formulation
\cite{Starobinskii} and it was shown that a gravity-driven
inflation can be achieved.

\textbf{3. Palatini formulation of $R^2$ gravity}

The field equations follow by substituting Eq.(\ref{R2}) into
Eq.(\ref{2.2})
\begin{equation}
(1+\frac{2R}{3\beta})R_{\mu\nu}-\frac{1}{2}g_{\mu\nu}(R+\frac{R^2}{3\beta})=\kappa^2
T_{\mu\nu}\ .\label{R2field}
\end{equation}

Contracting indices gives
\begin{equation}
R=-\kappa^2 T=\kappa^2\rho_m\ .\label{RR}
\end{equation}
The second equality follows because the radiation has vanishing
trace of momentum-energy tensor. This equation is quite
remarkable, since it is formally the same as the one given by GR,
with only one difference: $R_{\mu\nu}$ is associated with the
conformal transformed matric $h_{\mu\nu}=L'(R)g_{\mu\nu}$ and
$R=g^{\mu\nu}R_{\mu\nu}$.

From the conservation equation $\dot{\rho_m}+3H\rho_m=0$ and
Eq.(\ref{RR}), we can find that
\begin{equation}
\partial_0L'=-2\frac{\kappa^2\rho_m}{\beta}H\ .\label{}
\end{equation}

Substituting this into Eq.(\ref{aa}) we can get the Modified
Friedmann equation for the $R^2$ gravity:
\begin{equation}
H^2=\frac{2\kappa^2(\rho_m+\rho_r)+\frac{(\kappa^2\rho_m)^2}{3\beta}}{(1+\frac{2\kappa^2\rho_m}{3\beta})(6+3F_0(
\frac{\kappa^2\rho_m}{\beta})(1+\frac{1}{2}F_0(\frac{\kappa^2\rho_m}{\beta}))}\
,\label{R2MF}
\end{equation}
where the function $F_0$ is given by
\begin{equation}
F_0(x)=-\frac{2x}{1+\frac{2}{3}x}\ .\label{F0}
\end{equation}
It is interesting to see from Eq.(\ref{R2MF}) that all the effects
of the $R^2$ term are determined by $\rho_m$. If $\rho_m=0$,
Eq.(\ref{R2MF}) simply reduces to the standard Friedmann equation.

Now let's come to the discussion of inflation. To begin with, note
that in the metric formulation of the $R^2$ gravity, inflation is
driven by the vacuum gravitational field, i.e. we assume that the
radiation and matter energy densities is zero during inflation,
thus called "gravity-driven" inflation. However, in the Palatini
formulation, when the radiation and matter energy densities is
zero, it can be seen directly from Eq.(\ref{R2MF}) that the
expansion rate will be zero and thus no inflation will happen.
Thus, in the Palatini formulation of $R^2$ gravity, we cannot have
a gravity-driven inflation. So the only hope that the $R^2$ term
can drive an inflation without an inflaton field is that the
relationship between the expansion rate and the energy density of
radiation and matter will be changed which can lead to inflation
(thus what we are talking now is similar to the "Cardassian"
scenario of Freese and Lewis \cite{freese}: the current
accelerated expansion of the universe is driven by the changed
relationship between the expansion rate and matter energy
density). We will see that naturally there will be no inflation
and a power-law inflation can happen only under specific
assumption on $\rho_m$ and $\rho_r$.

First, in typical model of $R^2$ inflation, $\beta$ is often taken
to be the order of the Planck scale \cite{Starobinskii}. This is
also the most natural value of $\beta$ from an effective field
point of view. Thus we naturally have $\kappa^2\rho_m/\beta \ll
1$. Under this condition, it can be seen that from Eq.(\ref{F0}),
we have $F_0\sim 0$, and the MF equation (\ref{R2MF}) reduces to
the standard Friedmann equation:
\begin{equation}
H^2=\frac{\kappa^2}{3}(\rho_m+\rho_r)\ .\label{st}
\end{equation}
Thus it is obvious that in this case there will be no inflation.
Also note that from the BBN constraints on the Friedmann equation
\cite{Carroll4}, $\beta$ should be sufficiently large so that the
condition $\kappa^2\rho_m/\beta \ll 1$ is satisfied at least in
the era of BBN. Thus we conclude that in the most natural case,
Palatini formulation of $R^2$ gravity cannot lead to inflation.

Second, let's assume that in the very early universe, we have
$\kappa^2\rho_m/\beta \gg 1$. In this case, from Eq.(\ref{F0}),
the MF equation (\ref{R2MF}) will reduce to
\begin{equation}
H^2=\frac{\kappa^2\rho_m}{21}+\frac{2\beta\rho_r}{7\rho_m}+\frac{2\beta}{7}\
.\label{33}
\end{equation}
Then we can see that if the $\beta$ term could dominate over the
other two terms, it would drive an exponential expansion by the
effective cosmological constant $\beta$. But note that this
equation is derived under the assumption that $\beta\ll
\kappa^2\rho_m$. Thus inflation cannot be driven by the $\beta$
term. On the other hand, if we assume further that
$\rho_r\gg\kappa^2\rho_m^2/\beta$, i.e. the second term dominates
in the MF equation (\ref{33}), then from the relation
$\rho_r\propto a^{-4}$ and $\rho_m\propto a^{-3}$, the MF equation
(\ref{33}) can be solved to give $a\propto t^2$. Thus, only in
this case, we can get a mild power-law inflation. However, current
constraint on the rate of power-law inflation reads $p>21$ where
$a\propto t^p$ (see, e.g., Ref.\cite{lim}). So this case is not a
viable model of inflation.

\textbf{4. Conclusions and discussions}

In summary, in the Palatini formulation, the modified gravity
theory with a $R^2$ correction term would not lead to a
gravity-driven inflation, in opposite to the conclusion when
considering the theory in the metric formulation. And only under
the conditions that $\kappa^2\rho_m/\beta\gg 1$ and
$\rho_r\gg\kappa^2\rho_m^2/\beta$ we can get a power-law inflation
$a\propto t^2$. The difference of those two formulations is now
quite obvious. At present, we still can not tell which formulation
is physical. But this makes those results more interesting. It is
conceivable that quantum effects of the $R^2$ theory in Palatini
formulation would also be different from the metric formulation
(see Ref.\cite{Buchbinder} for a review). Such higher derivative
terms similar to the $R^2$ term may be induced by quantum effects
such as trace anomaly \cite{Buchbinder, Odintsov2}. It has been
recently shown \cite{Odintsov2} that phantom cosmology implemented
by trace anomaly induced terms also admits both early time
inflation and late time cosmic acceleration. It follows from our
consideration that $R^2$ term in Palatini formulation do not
support gravity-driven inflation, then we expect that also in
phantom cosmology with quantum effects in Palatini formulation,
the inflation does not occur.

There are many activities in the study of quantum versions of
$R^2$ gravity which seems to be a multiplatively renormalizable
theory (for a review, see Ref.\cite{Buchbinder}). However, such a
theory has had a serious problem: possible non-unitarity due to
the presence of higher derivative terms. It is very promising that
in Palatini formulation higher derivative terms do not play such a
role as in metric formulation. We expect that the unitarity
problem of $R^2$ gravity may be resolved in Palatini formulation.
This deserves further investigation.

Finally, it is interesting to explore the $R^2$ correction to the
chaotic inflation scenario \cite{R2} in Palatini formulation. When
written in Einstein frame and in metric formulation, this will
correspond to two scalar field inflation; in the Palatini
formulation, the model will correspond to a type of k-inflation
\cite{k-inflation}. More detailed investigations of this idea will
be interesting topic of future works.

\textbf{Acknowledgements}

We would especially like to thank Professor Sergei Odintsov for a
careful reading of the manuscript and many very helpful comments,
which have improved this paper greatly. Specially, he told us the
non-unitarity problem in $R^2$ gravity and suggested reconsidering
this problem in Palatini formulation. We would also like to thank
\'{E}anna Flanagan, Shin'ichi Nojiri for helpful correspondence
and Mauro Francaviglia and Igor Volovich for helping us finding
their earlier works. This work is partly supported by China NSF,
Doctoral Foundation of National Education Ministry and an
ICSC-World laboratory scholarship.

\begin{appendix}
\end{appendix}


\begin{thebibliography}{99}
\bibitem{Perlmutter} S. Perlmutter el al. Nature {\bf 404} (2000) 955;
Astroph. J. {\bf 517} (1999) 565; A. Riess et al. Astroph. J. {\bf
116} (1998) 1009; ibid. 560 (2001) 49; Y. Wang, Astroph. J. {\bf
536} (2000) 531; D.N.Spergel, et al., astro-ph/0302207; L.Page et
al. astro-ph/0302220; M.Nolta, et al, astro-ph/0305097; C.Bennett,
et al, astro-ph/0302209.
\bibitem{Carroll2}
S.M.Carroll, Living Rev. Rel. {\bf 4} (2001) 1; astro-ph/0310342
[astro-ph/0004075]; T. Padmanabhan, Phys. Rept. {\bf 380} (2003)
235 [hep-th/0212290].
\bibitem{Dark}
S.M.Carroll, M.Hoffman and M.Trodden, astro-ph/0301273; S.Nojiri
and S.D.Odintsov, hep-th/0303117; R. R. Caldwell, R. Dave and P.
J. Steinhardt, Phys. Rev. Lett. {80} (1998) 1582; A. Kamenshchik,
U. Moschella and V. Pasquier, Phys. Lett. B {511} (2001) 265; A.
Frolov, L. Kofman and A. Starobinsky, hep-th/0204187; T. Inagaki,
X. H. Meng and T. Murata, hep-ph/0306010; E. Elizalde, J. E.
Lidsey, S. Nojiri and S. D. Odintsov, hep-th/0307177.
\bibitem{Lue}
A. Lue, R. Scoccimarro and G. Starkman, astro-ph/0307034.
\bibitem{Dvali}
G. Dvali, G. Gabadadze and M. Porrati, Phys. Lett. B {\bf 485}
(2000) 208.
\bibitem{Carroll}
S.M.Carroll, V.Duvvuri, M.Trodden and M. Turner, astro-ph/0306438.
\bibitem{Capozziello}
S. Capozziello, S. Carloni and A. Troisi, astro-ph/0303041.
\bibitem{Odintsov3}
S. Nojiri and S. D. Odintsov, Phys. Lett. B {\bf 576} (2003) 5
[hep-th/0307071].
\bibitem{Dick}
R. Dick, gr-qc/0307052.
\bibitem{Woodard}
M. E. Soussa and R. P. Woodard, astro-ph/0308114.
\bibitem{Vollick}
D. N. Vollick, Phys. Rev. D {\bf 68} (2003) 063510
[astro-ph/0306630].
\bibitem{Volovich}
M.Ferraris, M.Francaviglia and I.Volovich, Nouvo Cim. B {\bf 108}
(1993) 1313 [gr-qc/9303007]; ibid, Class. Quant. Grav. {\bf 11}
(1994) 1505.
\bibitem{Flanagan}
\'{E}. \'{E}. Flanagan, Phys. Rev. Lett. {\bf 92} (2004)071101
[astro-ph/0308111].
\bibitem{Dolgov}
A. D. Dolgov and M. Kawasaki, Phys. Lett. B {\bf 573} (2003) 1
[astro-ph/0307285].
\bibitem{Wang}
X. H. Meng and P. Wang, Class. Quant. Grav. {\bf 21} (2004) 951
[astro-ph/0308031]; ibid, Class. Quant. Grav. {\bf 20} (2003) 4949
[astro-ph/0307354].
\bibitem{Chiba}
T. Chiba, Phys. Lett. B {\bf 575} (2003) 1 [astro-ph/0307338].
\bibitem{olmo}
G. J. Olmo and W. Komp, gr-qc/0403092.
\bibitem{Liddle}
A. R. Liddle and D. H. Lyth, Cosmological Inflation and Large
Scale Structure, Cambrigde University Press, 2000.
\bibitem{Starobinskii}
A. A. Starobinsky, Phys. Lett. B {\bf 91} (1980) 99.
\bibitem{Odintsov}
S. Nojiri and S. D. Odintsov, Phys. Rev. D {\bf 68} (2003) 123512
[hep-th/0307288].
\bibitem{Carroll4}
S.M.Carroll and M.Kaplinghat, astro-ph/0108002; K.A.Olive,
G.Steigman and T.P.Walker, Phys.Rept. {333-334} (2000) 389
[astro-ph/9905320]; S.Burles, K.M.Nollett, J.N.Truran and
M.S.Turner, Phys.Rev.Lett. {82} (1999) 4176 [astro-ph/9901157];
D.Tytler, J.M.O'Meara, N.Suzuki and D.Lubin, astro-ph/0001318.
\bibitem{Magnano}
G.Magnano and L.M.Sokolowski, Phys. Rev. D {\bf 50} (1994) 5039
[gr-qc/9312008].
\bibitem{Buchbinder}
I.L.Buchbinder, S.D.Odintsov and I.L.Shapiro, Effective Action in
Quantum Gravity, IOP Publishing, 1992.
\bibitem{Odintsov2}
S.Nojiri and S.D.Odintsov, hep-th/0308176.
\bibitem{R2}
S.Gottlober, V.Muller and A.Starobinsky, Phys.Rev.D43 (1991) 2510;
V.H.Cardenas, S. del Campo and Ramon Herrera, gr-qc/0308040; X. H.
Meng and P. Wang, Class. Quant. Grav., {\bf 21} (2004) 2029
[gr-qc/0402011].
\bibitem{k-inflation}
C. Armendariz-Picon, T. Damour and V. Mukhanov,
Phys. Lett. B {\bf 458} (1999) 209.
\bibitem{freese}
K. Freese and M. Lewis, Phys. Lett. B {\bf 540} (2002) 1
[astro-ph/0201229].

\bibitem{lim}
C. Armendariz-Picon and E. A. Lim, JCAP {\bf0312} (2003) 006
[hep-th/0303103].
\end{thebibliography}
\end{document}